\documentclass[12pt]{iopart}
\usepackage{siunitx}
\usepackage{array}
\usepackage{booktabs}
\usepackage{graphicx}% Include figure files
\usepackage{dcolumn}% Align table columns on decimal point
\usepackage{bm}% bold math
\usepackage[utf8]{inputenc}
\usepackage[T1]{fontenc}
\usepackage{mathptmx}
\usepackage{xcolor}

\begin{document}

\title[]{Remote multi-user control of the production of Bose-Einstein condensates for research and education}

\author{J S Laustsen, R Heck, O El{\'i}asson, J J Arlt, J F Sherson and C A Weidner}

\address{ 
Department of Physics and Astronomy, Aarhus University, 8000 Aarhus C, Denmark
}%

\ead{sherson@phys.au.dk}
\vspace{10pt}
\begin{indented}
\item[]\today
\end{indented}

\begin{abstract}
Remote control of experimental systems allows for improved collaboration between research groups as well as unique remote educational opportunities accessible by students and citizen scientists. Here, we describe an experiment for the production and investigation of ultracold quantum gases capable of asynchronous remote control by multiple remote users.
This is enabled by a queuing system coupled to an interface that can be modified to suit the user, e.g. a gamified interface for use by the general public or a scripted interface for an expert.
To demonstrate this, the laboratory was opened to remote experts and the general public. During the available time, remote users were given the task of optimising the production of a Bose-Einstein condensate (BEC). This work thus provides a stepping stone towards the exploration and realisation of more advanced physical models by remote experts, students and citizen scientists alike.
\end{abstract}
% Uncomment for keywords
\vspace{2pc}

\section{Introduction}
Ultracold quantum gases have become one of the prime platforms for simulating technologically relevant quantum systems within the last decades. In particular, extremely clean and pure quantum model systems can be realised that offer a high degree of controllability with respect to parameters such as the atoms' interaction strength and temperature. This progress has led to increasingly poweful and complex experiments in lattice-based quantum simulation~\cite{grossQuantumSimulationsUltracold2017, schaferToolsQuantumSimulation2020}, the simulation of strongly-correlated condensed matter systems~\cite{hofstetterQuantumSimulationStrongly2018}, and quantum computing with Rydberg atoms~\cite{saffmanQuantumComputingAtomic2016}, among others, rendering cold atoms a fantastic platform for the development of technologies that will drive the second quantum revolution~\cite{deutschHarnessingPowerSecond2020}.

Collaboration between experimental and theoretical groups is an essential part of developing and evaluating models applicable to quantum simulation experiments. To optimise experimental procedures, it may indeed be beneficial to use dedicated, automated protocols developed by theory groups. Opening the laboratory to direct remote control by collaborators may thus increase the efficiency of such collaborative efforts.
Moreover, a remote control system opens up new possibilities regarding outreach to students and the general public. By allowing a broad audience of non-expert users to control some experimental parameters, one can imagine a number of scenarios geared towards public outreach and education. First, the public can take part in citizen science experiments, and, in particular, previous work using the system described here shows that valuable insight into cognitive science can be gained~\cite{Heck2018}. Secondly such platforms can be used to educate and engage non-experts in quantum physics, e.g. by allowing students access to cutting-edge research laboratories regardless of where the laboratory is physically located. In both cases this creates the need for an intuitive user interface which allows users to focus on the essential parts and hides the technical details. At the same time, the experimental system must also contain the infrastructure to handle the input from one of many users and return the relevant results to the correct user. A number of these open platforms already exist, including the IBM Quantum Experience~\cite{IBM_quantum_experience}, and the open availability of this platform has allowed for the production of a number of research articles (see, e.g., Refs.~\cite{bertaEntropicUncertaintyMeasurement2016, dumitrescuCloudQuantumComputing2018, wang16qubitIBMUniversal2018, woottonRepetitionCode152018, zulehnerEfficientMethodologyMapping2019}), educational material~\cite{woottonTeachingQuantumComputing2020}, and games~\cite{decodoku}.

In principle, any experimental control program can be modified for remote control via the addition of a remote server and a suitable front-end for the user.
In terms of experimental control programs, several publicly available systems for cold atom experiments have been published~\cite{OwenFastLinebasedExperiment2004,GaskellOpensourceExtensibleSystem2009,StarkeyScriptedControlSystem2013,KeshetDistributedGUIbasedComputer2013,PeregoScalableHardwareSoftware2018}. In addition, numerous commercial options are available, such as systems by ColdQuanta~\cite{ColdQuantaControlSystem}, MLab~\cite{MLabsARTIQ}, ADwin~\cite{ADwin} and IQ Technologies~\cite{IQTechnologiesLaboratoryControlSystem} that can be purchased together with suitable hardware. All of these control systems have sub-microsecond control of digital and analog channels and some allow for direct output of radio frequency (RF) signals. Additionally, they typically allow for communication with external hardware through different protocols or via implementation of appropriate drivers. These criteria define the typical minimum viable product for useful cold-atom experiment control. Software for camera control and analysis of the images enables some systems to optimise experimental performance in a closed loop optimisation of experimental parameters. Moreover, all of these systems are remotely-controllable either directly or via simple screen-sharing protocols. However, to our knowledge, none of these control programs had been used in a multi-user setting where several users simultaneously remotely controlled an experiment through the use of the aforementioned server and front-end, with the exception that, while preparing this manuscript, we became aware of the Albert system built by ColdQuanta that came online in late 2020~\cite{CQ_Albert}. This work thus represents the growing commercial and academic interest in remote control of cold atom systems.

Here we discuss the implementation of a remote controlled experiment usable by single expert user or multiple non-expert users accessing the experiment. Previously, we have documented the challenges that we provided to our users, as well as the main findings that arose from this work~\cite{Heck2018}. However, we have not explained the underlying system architecture and the overarching possibilities that this gives rise to in research and education. The general knowledge of these details is crucial for other groups to implement similar systems, and this is what we focus on in this work. In both of the use-cases considered here, there is a need for a queuing system for the input sequences and the return of the results. When considering multi-user access there is also the need to track the sender throughout the process of queuing, performing the experiment, analysis and reporting the results. The infrastructure of the experiment also allows for multiple expert users and this option will be explored in future work. For instance, one could imagine running multiple collaborative efforts simultaneously.

This paper is organised as follows: In the first section the software enabling remote control is presented. Following this, the experimental sequence and its technical implementation is described. We then describe the two different implementations of remote user access that were used in previous work: single- and multi-user control~\cite{Heck2018}. The last section concludes and provides an outlook.

\section{The control software}

The experimental control system is LabVIEW-based and capable of being expanded as new hardware is added to the experiment. A field-programmable gate array (FPGA, PCI-7813R) is used to control $70$ digital and $48$ analog channels through $4$ digital to analog converter modules (NI 9264, NI 9263). In addition, the system can communicate with hardware drivers to other hardware such as motion stages, piezoelectric devices and RF synthesisers. Thus, our control system meets the aforementioned criteria for usability in a cold atom experiment.

The control program is based on a server/client architecture. The server controls all hardware, including the FPGA, and the client provides an interface for the user and compiles the programmed sequence.  On the client side the sequence is built of \textit{waves} which correspond to the output of a given digital or analog channel, a GPIB command or a command through a hardware driver.  Regularly-used sequences of waves can be collected and collapsed into \textit{blocks}, e.g., the commands required to load atoms into a trap or image an atom cloud. For each wave and block, externally-accessible variables can be declared, e.g. the frequencies of the RF tones applied to acousto-optic modulators (AOMs) or the duration of the RF pulse applied to the AOM. This allows the user to create sequences with an adaptive level of abstraction. For instance one can hide the exact technical implementation of experimental steps in a block but keep the essential control parameters accessible, which is useful for reducing the cognitive load of a remote user.

\begin{figure}[h!] %
			\includegraphics[width=\columnwidth]{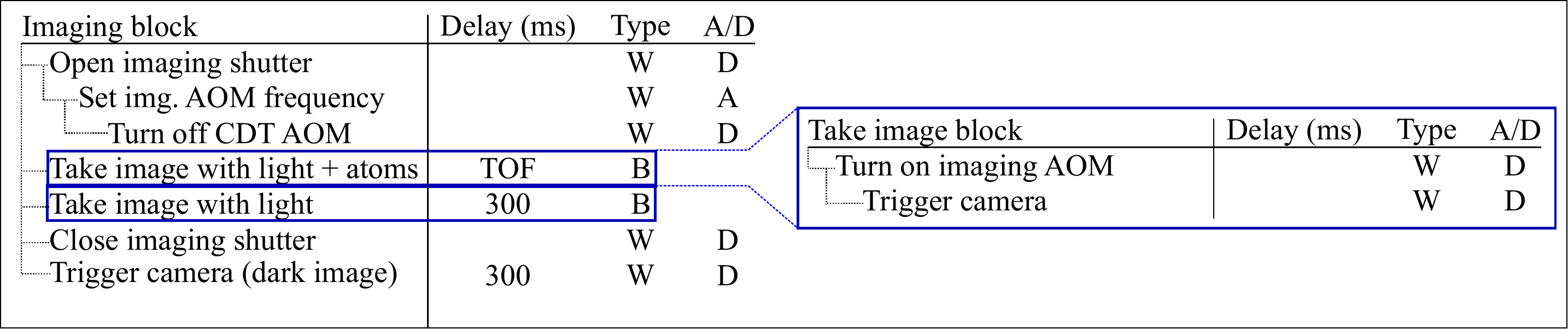} %
			\caption{An example of a block used take an absorption image at the end of an experimental sequence (cf. Sec.~\ref{sec:expt}). The block contains individual analog (A) and digital (D) waves (W), as well as two embedded blocks (B) used to take the absorption and background images. The block runs from the top to bottom with indented elements running in parallel with the element above. In this block, the imaging shutter is opened while the frequency applied to the imaging AOM is set and the CDT AOM is turned off (thus turning off the CDT itself and dropping the atoms from the trap). Then, after a variable time-of-flight (TOF) the next block (blue squares, with zoom-in to the right of the main block) simultaneously pulses the imaging AOM and triggers the camera shutter to take the absorption image of the atoms. After 300 ms of camera processing time, the same block takes a background image without atoms. The imaging shutter is closed, and after an additional 300 ms, the camera is triggered again to take the dark image without any light present. Subtracting the absorption image from the background and dark images reveals the atom signal.} %
			\label{fig:block} %
	\end{figure}

An example of a block used for absorption imaging of ultracold atoms is shown in Fig. \ref{fig:block}, where smaller blocks are incorporated. The waves and blocks are ordered in a tree structure that controls the timing of an experiment. The tasks are performed from top to bottom in such a tree. Any waves or blocks on indented branches are performed simultaneously, and delays can be defined within individual elements for more precise control of relative timing. Initialised outputs may be defined such that they either hold their last value or are reset to a default value after a given time. Thus the user need only handle the values of relevant outputs at any given point. Wave and block variables can be scanned individually or jointly in single- and multi-dimensional scans, respectively. Loops are also available where a subset of blocks is repeated while one or several variables are changed. For example, the user can loop the capture of atoms in a trap while changing a given parameter value during each loop iteration, effectively performing a parameter scan within a single realisation of the experiment.

The novel aspects of the control system lie in its capability for communication with remote users. This includes loading sequences from a queue either created by a single user or multiple different users. After a remotely-requested sequence is performed, relevant results (e.g. atom number) are sent back to the user who designed the sequence. To make the remote control as flexible as possible, the control software does not provide any user interface for the remote user but communicates with stand-alone interfaces. Thus a remote user can easily set up closed-loop optimisation by linking the returned results into a script running a given optimisation algorithm that then generates the next desired sequence, as described in detail below.

\section{\label{sec:expt}The experiment}

To demonstrate the use of the control system and the communication necessary for multi-user operation, we conducted an experiment in which remote users create a Bose-Einstein Condensate (BEC). The experimental system is described in Refs.~\cite{Bason2018,Eliasson2019} and only its main features are described here.

The experimental sequence starts by precooling a cloud of Rb-87 atoms in a 3D MOT. Here the atoms are laser-cooled and trapped via a combination of light pressure and magnetic field gradients. Subsequently polarisation gradient cooling is performed and the atoms are optically pumped to the low-field-seeking $|F \, , \, m_F\rangle = |2,2\rangle$ state. The atoms are then trapped in a magnetic quadrupole trap generated by a pair of coils in an anti-Helmholtz configuration. These coils are mounted on a rail and are used to transport the atoms through a differential pumping stage to the final chamber. Here the atoms are evaporatively cooled by transferring the hottest atoms to a high-field-seeking sublevel. By the end of the evaporation sequence the atoms have a temperature of roughly \SI{30}{\micro \kelvin}. Subsequently, a crossed optical dipole trap (CDT) consisting of two laser beams (wavelength $\lambda = \SI{1064}{\nano \meter}$, $1/e ^2$ waists of \SI{45}{\micro \meter} and \SI{85}{\micro \meter}) is superimposed on the atoms. After the final evaporation stage, the atom cloud is released from the trap and an absorption image is recorded after a TOF. If the user-defined evaporation sequence is effective, the cloud is condensed and a BEC is visible in the image.

For the remote experiments reported here, the control parameters available to the users are the laser powers of both laser beams forming the CDT and the current in the quadrupole coils as a function of time. This configuration allows the user to cool the atoms using forced evaporative cooling either in a pure CDT~\cite{grimm_optical_2000-1}, in a so-called hybrid trap consisting of the quadrupole magnetic field and one of the dipole beams~\cite{Lin2009}, or any combination of the two. The depth and geometry of the trap depends on these parameters in a non-trivial optimal way, providing an opportunity for external optimisation, the goal of which is to produce the largest possible BEC. For both expert and non-expert users a limitation of the available control space is necessary as only a small fraction of the full control landscape will yield a BEC.

%%%{Remote control implementation}
\section{Two cases of remote user control}

For a remotely-controllable system to be useful, appropriate user interfaces must be developed, and each user class has different requirements to optimally facilitate the interaction. For experts a scripted interface can be an advantage as complex algorithms can be directly implemented. A more visual interface of the control (for instance in a game-like setting) is needed for non-expert users. Importantly, a different program structure is needed when handling input from a single user or multiple users. In what follows, we describe two different implementations of our remote control geared towards single expert users and asynchronous use among the general public, respectively. In this section, we elaborate on each of these cases. Again, note that the data presented here is drawn from the same source as our initial work~\cite{Heck2018}, and detailed research results can be found there. Here, we focus on more technical aspects of the experimental implementation and execution.

\subsection{Single-user remote control}

% see Z:\experiment\DATA_ALICE - clean\2016\07\2016-07-04\dir_input_test_ramps_160704 for details on how the input ramps looked
In our first implementation of remote control, an expert user optimised the evaporation using the so-called dressed chopped random-basis (dCRAB) optimisation algorithm~\cite{Rach2015}. Note that the algorithm was implemented on the user side, so our implementation of remote expert control is algorithm agnostic. Here the user had access to the CDT laser powers and quadrupole coil currents as a function of time. Sequences of waveforms corresponding to the parameter values were created as text files and sent to the experimental control program through a folder on a cloud drive and placed in a queue. Even for a single user, a queue is necessary due to the relatively long ($30$~s) cycle time of the experiment. The queue operated on the first-in-first-out (FIFO) principle, allowing the user to submit several sequences simultaneously and easily keep track of the outputs; this is useful, e.g., when initialising the initial simplex for Nelder-Mead optimisation.

For each user-accessible parameter, the parameter values can be defined at any desired time, while values between these times are linearly interpolated at the hardware level. Therefore the effective temporal resolution of the waveforms can be controlled by the user, and the total number of parameter/time pairs that can be used is ultimately limited by the memory of our FPGA.
% These waveforms then set the time-dependent control variables corresponding to the experimental parameters accessible by the user.

When a given sequence was ready to be run, the relevant experimental sequences were generated by reading in the waveforms the from text files generated by the expert user. The experiment was then run and the resulting image was analysed. From this image the BEC atom number was extracted and returned to the expert user through the same cloud drive, again as a text file. This atom number was read in by the expert user and served as the cost parameter closing the optimisation loop.

\subsection{Multi-user remote control}

\begin{figure}[] %
			\includegraphics[width=\columnwidth]{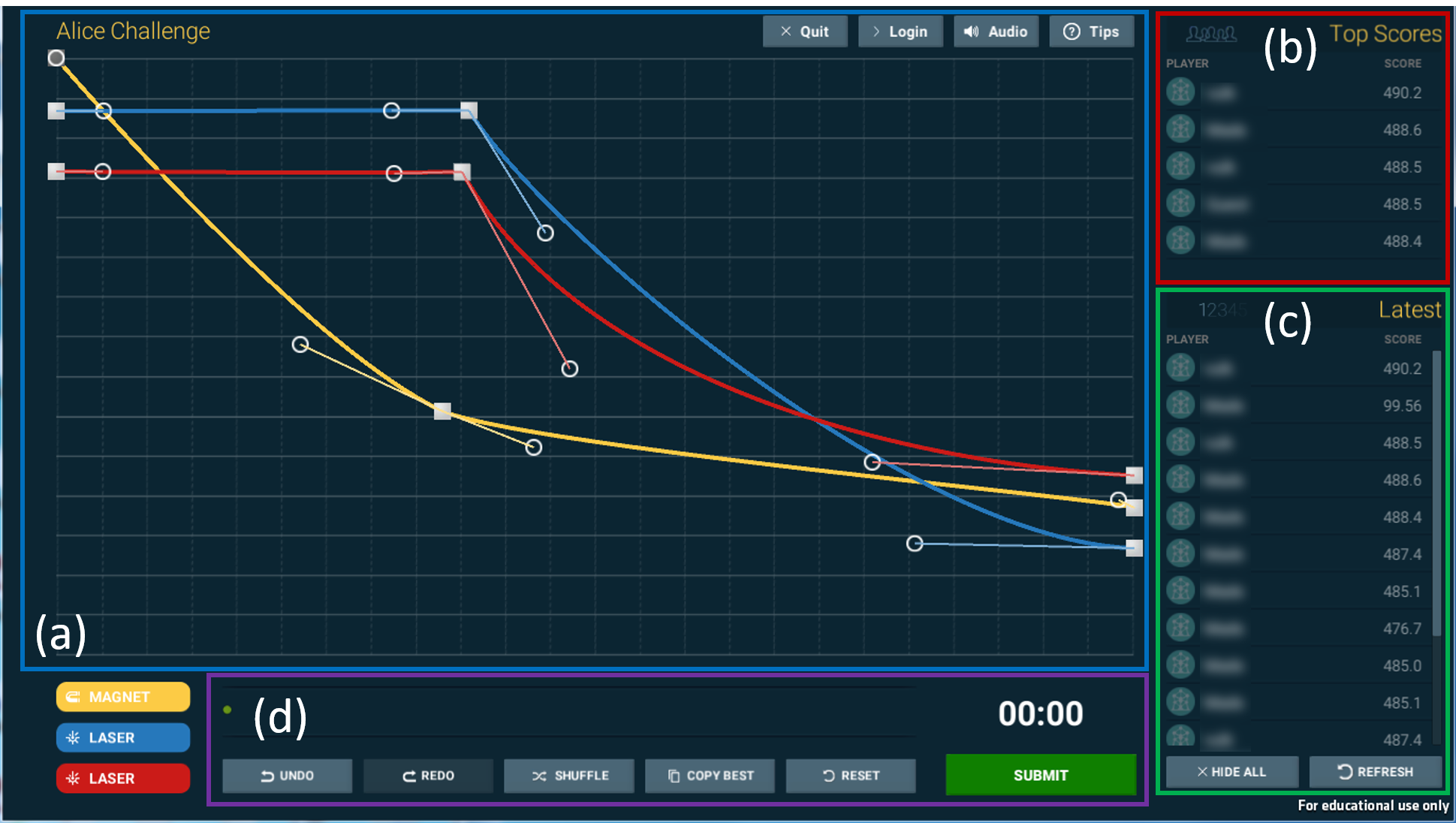} %
			\caption{A screenshot of the interface used in the \emph{Alice Challenge}, showing (a) the spline editor used to create the ramps of the laser powers and coil current, shown on a logarithmic scale, (b) top score list, (c) latest executed sequences, and (d) the control buttons, including the estimated wait time until the submitted sequence is returned.}%
			\label{fig:screenshot}
	\end{figure} %

In the second implementation, called the \textit{Alice Challenge}, citizen scientists were given access to the experiment. This subsection details the architectural considerations required for the challenge as well as some statistics on user load in real time over the course of the challenge. This information is useful when considering the future implementation of similar systems.

Citizen scientists were given access to the system via a gamified interface as shown in Fig. \ref{fig:screenshot}, and this is used to provide more intuitive access to the parameter space. The interface was designed to visualise the ramps of the laser powers and coil currents sent by the citizen scientists to the experiment. The control values were normalised and presented for ease of use on a logarithmic axis in a spline editor where the user could manipulate the curves by clicking-and-dragging points along the curve. When the user was done editing the curves, the sequence was submitted and subsequently realised in the experiment.

This was done in the following manner: The user sequence (encoded as a JSON file with a unique user ID) was delivered to a web server. The web server then delivered the sequence to the cloud folder that served as the queue. When a sequence was ready to be evaluated, it was sent to the LabVIEW control system, where the JSON data was translated into waveform data identical to the type used in the single expert user configuration. This was done via a special \emph{optimisation class} defined in LabVIEW that was responsible for extracting the relevant parameters from the JSON file. Once a sequence was completed, the control program wrote the results to another JSON file, inserted the relevant user ID, and stored it in a separate folder on the cloud. % user generates sequence in-game and submits it:
% sequence becomes JSON (only for multi-user) and is delivered via web server to the Dropbox queue
% when a sequence is ready to be evaluated:
% Alice translates JSON into waveform data (that looks like the single-user waveform data)
% JSON also contains an "object ID" which tracks the users (which user did what)
% optimisation class in LV is taking the relevant parameters, compiling, and sending the proper sequence
% Alice writes results to JSON (keeping track of object ID) and then this is sent back to the user to be displayed in the game interface
% the results are delivered back through the web server
% game interface scales the score, etc
% NB: queue and results are generated in different folders of the same Dropbox
The webserver then delivered the results, and the backend of the game interface scaled the BEC atom number to a \emph{score} which was displayed to the user. The score and corresponding sequence was also visible for other players who could then copy the sequence as inspiration when creating their own sequences.

\begin{figure*}[] %
			\includegraphics[width=\textwidth]{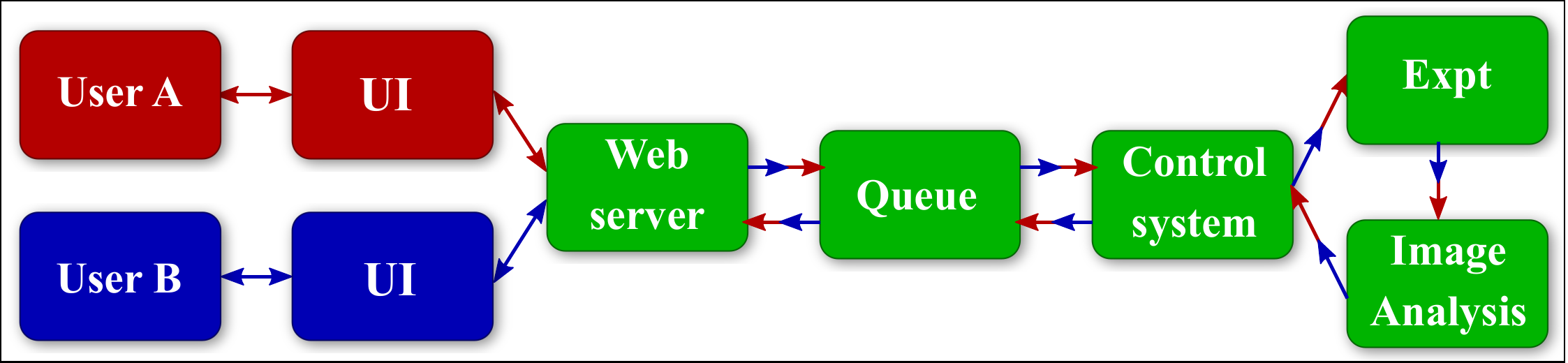} %
			\caption{
			Schematic view of the data flow for the remote control of the experiment by multiple, asynchronous outside users during the \emph{Alice Challenge}. Experimental sequences are submitted through a game-like user interface to a web server that subsequently sends them to the cloud-based queue in the order they are received (here, User A has submitted their sequence first). Each submission has a unique User ID that is tracked throughout the process. The control system reads the oldest files via the FIFO principle and runs the corresponding experiment. When image analysis is completed, the results are returned to the proper user via the UI.
			}
			\label{fig:schematic} %
\end{figure*} %

In contrast to the case of a single user, the web server was needed to track the run number and user ID if multiple users were running the experiment simultaneously. A schematic view of the multi-user data handling infrastructure used for the \textit{Alice Challenge} is presented in Fig. \ref{fig:schematic}. To ensure that the result of the experiment was linked to the right user sequence a check was made in the experimental control system such that the experimental sequence was repeated if no result was returned for a given run.

Moreover, the state of the experiment was checked by inserting an established benchmark sequence in the queue every tenth run. This benchmark sequence was known to create a BEC under stable experimental conditions. In the case of a problem, such as a laser failure, the experiment was paused until the problem was solved. At the same time, the users were informed of the temporary delay caused by the disturbance. The benchmark sequence was also executed in case of an empty queue in order to keep the experiment in a stable condition. This also allows one to track overall experimental drifts, e.g. due to thermal fluctuations, which can be useful in advanced closed-loop optimisation schemes.

\begin{figure}[] %
		\includegraphics[width=0.8\columnwidth]{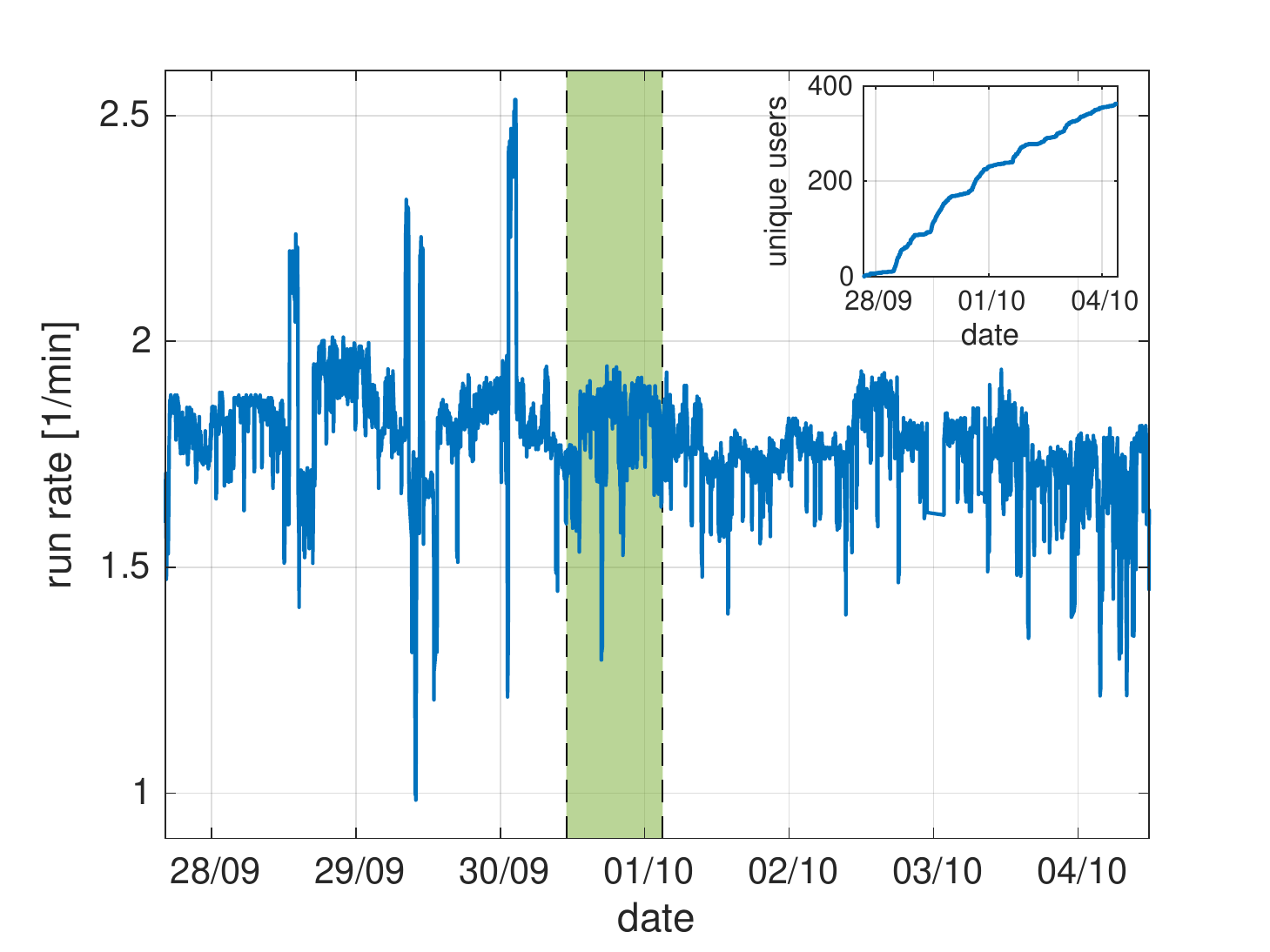}\raggedleft %
		\caption{The mean rate at which the experimental sequences were performed during the week in which the experiment was open to non-expert outside users. The different plateaus arise due to changes in the ramp time whereas the high run rates are an effect of synchronisation problems. The green shading denotes the time period depicted in Fig. \ref{fig:queue}. In the inset the accumulated number of unique citizen scientists that used the system throughout the week is shown. The date markers indicate midnight CET on a given day.} %
		\label{fig:runrate} %
\end{figure} %

The experiment was open to the public for a full week with only minor interruptions, resulting in total of $7577$ submitted and evaluated sequences. Figure \ref{fig:runrate} shows the rate of the experimental runs during this week. Over the course of the challenge, the preset duration of the evaporation ramps were varied. This allowed the citizen scientists to explore different optimisation landscapes, varying the challenge offered to them and keeping things interesting for returning users. 

The different evaporation durations create some variation of the rate at which experiments were performed over the course of the week. In addition, when experimental problems caused the experiment to be paused, the rate decreased. It should also be noted that the peaks of high run rates were caused by synchronisation problems between the web server and the control program. This problem was solved on the third day of the challenge, after which none of the larger peaks are visible. The inset shows the progress of the accumulated number of unique users through out the week.

\begin{figure}[] %
		\includegraphics[width=0.8\columnwidth]{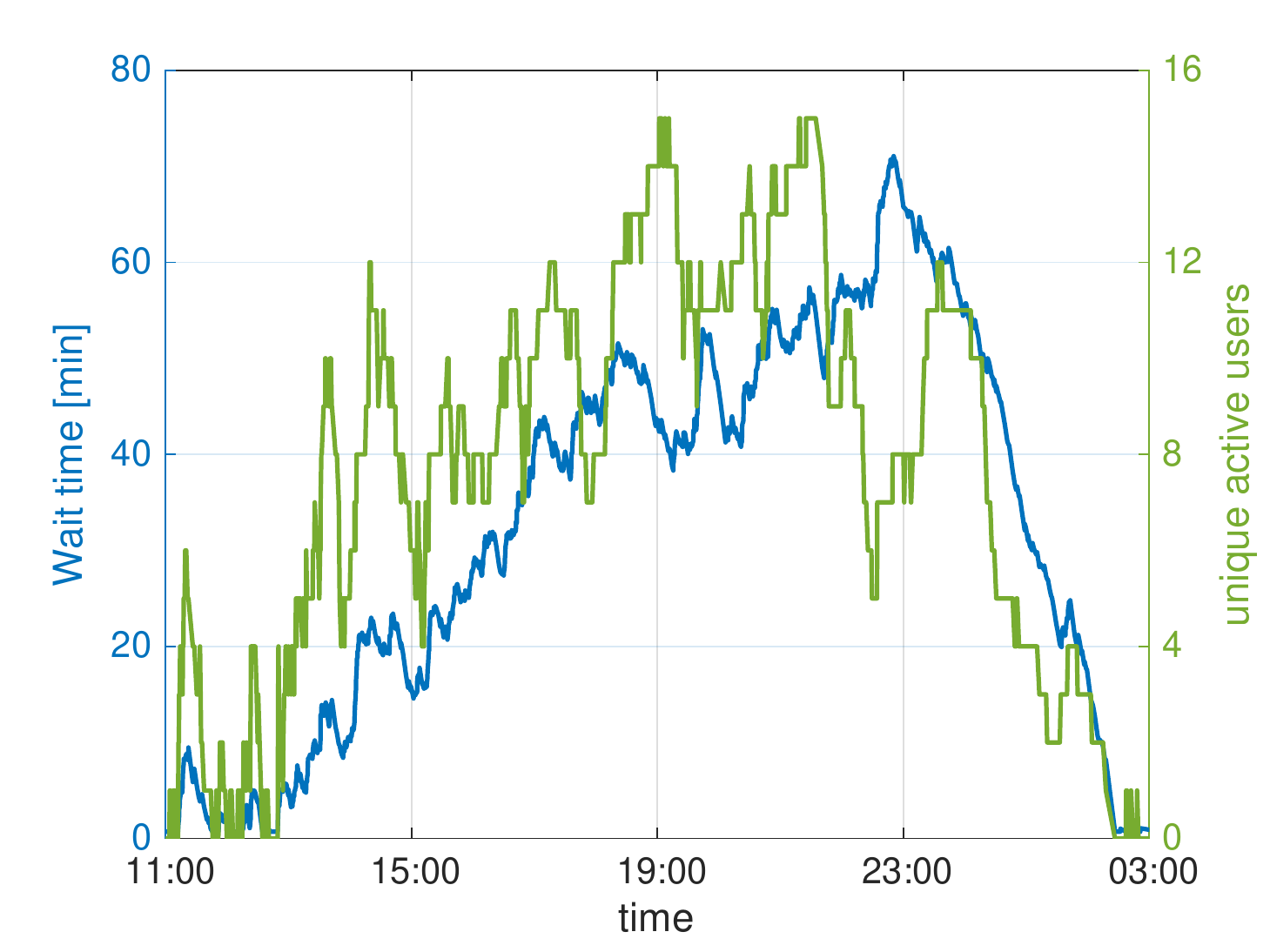}\raggedleft %
		\caption{The queue time and number of active users during the busiest period of the challenge. The blue trace shows the calculated waiting time from a submission of a sequence until a result is given and the green trace shows the number of active users at any given point.} %
		\label{fig:queue} %
\end{figure}

Throughout the day the number of active users varied as players from several parts of the world came online. Figure \ref{fig:queue} shows the queue and number of active users on Friday evening (CET), one of the highest peaks in active users. Here we see that up to 15 unique users were active at any given time during the evening which created a wait time of above one hour. As the number of users declined, the length of the queue was slowly reduced. Since each user could submit several different sequences at a time, the correlation between the number of unique users and the queue length is nonlinear.

\begin{figure}[] %
		\includegraphics[width=0.8\columnwidth]{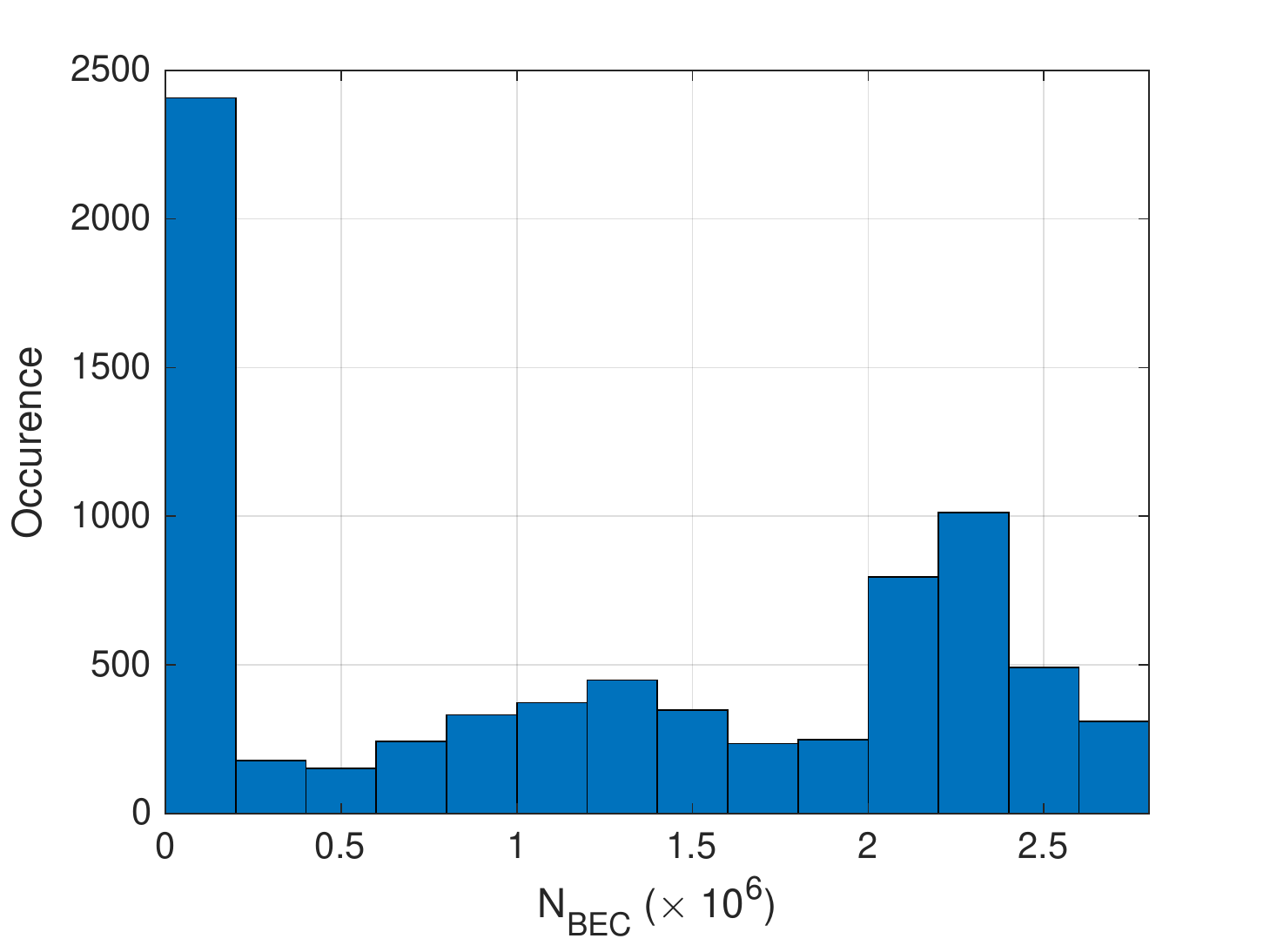}\raggedleft %
		\caption{A histogram of how many times a given BEC atom number was obtained by the sequences submitted by the users of the Alice challenge. Above $73\%$ of the submitted sequences created a BEC. This data is also presented in Ref.~\cite{Heck2018}.} %
		\label{fig:histogram} %
\end{figure}

Figure \ref{fig:histogram} shows a histogram of how many times a given BEC number was achieved. We see that most sequences submitted to the experiment result in the creation of a BEC. This is despite the fact that citizen scientists had limited insight to the physical system they were controlling to create the condensates. 

\section{Outlook}

In future work, remote controlled optimisation of a system may be advantageous, as remote optimisation allows for easy implementation of advanced optimisation algorithms. Several programs are available that can implement closed-loop optimisation of cold-atom experiments~\cite{OwenFastLinebasedExperiment2004,GaskellOpensourceExtensibleSystem2009,StarkeyScriptedControlSystem2013,KeshetDistributedGUIbasedComputer2013,PeregoScalableHardwareSoftware2018}. Students can also access such systems for educational purposes, as has already been done with quantum computers~\cite{woottonTeachingQuantumComputing2020}. This allows students to explore complex, cutting-edge research systems that are not accessible in many educational learning laboratories. 

For remote users to be able to run optimise the experiment, the relevant experimental control parameters have to be easily controllable. Collaborative optimisation between several remote users requires a structure that includes a multi-user queue and tracking the ID of submitted sequences so that the results may be returned to the correct user. The control program presented here can be expanded to give remote access to larger parts of the control sequence or even the entire experiment. Thus, future work will give remote users expanded access, allowing them to tackle more advanced scientific problems in a research or educational setting.  For example, with the new capabilities of the experiment to image single atoms using a quantum gas microscope~\cite{eliassonSpatialTomographyIndividual2020} in combination with spin addressing~\cite{Weitenberg2011b} and arbitrary light potential generation techniques~\cite{ChristieS.ChiuQuantumStateEngineering2018} the experiment can be used as an analog quantum simulator with remote control capability. 

Such advanced control will require a more complex user interface since the number of experimental parameters would increase, rendering algorithmic optimisation more difficult. However other groups have shown that optimisation of such systems with large parameter spaces is possible using genetic algorithms~\cite{Rohringer2011,Lausch2016,Geisel2013} or machine learning methods such as neural networks~\cite{tranterMultiparameterOptimisationMagnetooptical2018a}, Gaussian processes~\cite{Wigley2016a} or evolutionary optimisation~\cite{barkerApplyingMachineLearning2020}.

\section{Acknowledgements}
The authors would like to acknowledge Aske Thorsen for the development of the LabVIEW control code used for the experiments described here.

\section*{References}
\bibliographystyle{iopart-num}
\bibliography{AlicePaperEJP} 

\end{document}